\documentclass[preprint,aps]{revtex4}

\usepackage{graphicx}
\usepackage{dcolumn}
\usepackage{bm}

\newcommand{\Vckm}{V_{\text{CKM}}}
\newcommand{\Vpmns}{V_{\text{PMNS}}}
\newcommand{\TVpmns}{\tilde{V}_{\text{PMNS}}}

\newcommand{\delsp}{\delta_{\text{spon}}}

\begin{document}

\title{Spontaneous CP Violating Phase as the Phase in PMNS Matrix}

\author{Xiao-Gang He$^{1,2}$, and
Lu-Hsing Tsai$^2$}
\affiliation{
$^1$INPAC, Department of Physics, Shanghai Jiao Tong University, Shanghai\\
$^2$Department of Physics and Center for Theoretical Sciences, National Taiwan University, Taipei\\
}

\date{\today} 

\begin{abstract}
We study the possibility of identifying the CP violating phases in the PMNS mixing matrix in the lepton sector and also that in the CKM mixing matrix in the quark sector with the phase
responsible for the spontaneous CP violation in the Higgs potential, and some implications. Since the phase in the CKM mixing matrix is determined by experimental data, the phase in the lepton sector is therefore also fixed. The mass matrix for neutrinos is constrained leading to constraints on the Jarlskog CP violating parameter $J$, and the effective mass $<m_{\beta\beta}>$ for neutrinoless double beta decay. The Yukawa couplings are also constrained. Different ways of identifying the phases have different predictions for $\mu \to e e\bar e$ and $\tau \to l_1 l_2  \bar l_3$. Future experimental data can be used to distinguish different models.

\end{abstract}

\maketitle

\newpage

\section{Introduction}

The origin of CP violation is one of the most outstanding problems of
modern particle physics. In the quark sector, the leading effect of CP violation comes from the
phase $\delta_{\text{KM}}$, which sometimes is referred as the Dirac phase, in the Cabbibo-Kobayashi-Maskawa (CKM)
model~\cite{cab,km}. CP violation may also exist in the
lepton sector. One of the ways to describe low energy CP violation in this sector is to have CP violating Dirac phase
$\delta_{\text{PMNS}}$ in the Pontcove-Mkai-Nagawa-Sakata (PMNS) mixing matrix~\cite{Pontecorvo:1957cp,Maki:1962mu}.
If neutrinos are Majorana particles, besides the Dirac phase $\delta_{\text{PMNS}}$,
there may also exist CP violating Majorana phases. It is important to understand the origin of CP violation.
An interesting proposal due to T.-D. Lee is that CP is spontaneously
violated~\cite{tdlee}. The popular Weinberg model~\cite{weinberg1}
of spontaneous CP violation model has problems~\cite{problem,bigi}
with data and has been decisively ruled out by CP violating
measurement in B decays~\cite{pdg2010}. In a previous work, we have studied the possibility of restoring the idea that CP
is broken spontaneously and identifying the phase $\delta_{\text{KM}}$, up to a sign, as
the phase $\delsp$ that causes spontaneous CP violation in
the Higgs potential~\cite{Chen:2007nx}. In this work we generalize this approach to the lepton sector to study how CP violating phase $\delta_{\text{PMNS}}$
may be identified with the phase $\delta_{\text{spon}}$, and some of the consequences.

A model to realize the above idea for quark sector has been studied in Ref.~\cite{Chen:2007nx}. It involves multi-Higgs fields.
In that work, in order to avoid possible strong CP problem a Peccei-Quinn symmetry~\cite{pq} is imposed to the model~\cite{bigi}.
Implementing spontaneous CP violation with PQ symmetry
then requires more than two Higgs doublets~\cite{hvgn}. For our purpose we find that in order to have
spontaneous CP violation with PQ symmetry at least three Higgs
doublets $\phi_i = e^{i\theta_i}H_i = e^{i\theta_i}(
{1\over \sqrt{2}}(v_i +R_i +i A_i),
h^-_i)^T$ and one complex Higgs singlet
$\tilde S = e^{i\theta_s}S = e^{i\theta_s}(v_s + R_s + i
A_s)/\sqrt{2}$ are needed. The Higgs singlet with a large vacuum
expectation value (VEV) renders the axion from PQ symmetry
breaking to be invisible~\cite{invisible,kk}, thus satisfying
experimental constraints on axion couplings to fermions.

Two possible ways of assigning PQ charges to the quarks and Higgs bosons were considered~\cite{Chen:2007nx},
\begin{eqnarray}
\mbox{Model a)}&&Q_L : 0\;,\;\;U_R: -1\;,\;\;D_R:
-1\;,\;\;\phi_{1,2}:
+1\;,\;\;\phi_d=\phi_3: -1;\nonumber\\
\mbox{Model b)}&&Q_L : 0\;,\;\;U_R: +1\;,\;\;D_R:
+1\;,\;\;\phi_{1,2}: +1\;,\;\;\phi_u=\phi_3: -1\;.\label{quarkPQ}
\end{eqnarray}
In both cases, $\tilde S$ has PQ charge $+2$.

With these PQ charge assignments, it is possible to have spontaneous CP violation with a
non-zero phase of VEVs, $\delsp = \theta_2 - \theta_1$.
The other phases $\theta_3$ and $\theta_s$ are related to $\delsp$ by minimization conditions.

The resulting quark mass terms in the Lagrangian, after removing un-physical phases and the diagonalized basis of down quark mass matrix $\hat M_d$
and up quark mass matrix $\hat M_u$ for models a) and b), respectively, we have~\cite{Chen:2007nx}
\begin{eqnarray}
\mbox{Model a)}: \mathcal{L}_m &= &-\overline{ U_L} \left [M_{u1}+ M_{u2} e^{i\delsp}\right ]U_R
- \overline{ D_L} \hat M_d D_R + h.c.\;,\nonumber\\
\mbox{Model b)}: \mathcal{L}_m &= &-\overline{ U_L} \hat M_u U_R
- \overline{ D_L}\left [M_{d1}+ M_{d2} e^{-i\delsp}\right ]D_R + h.c.\;,\nonumber\\
\end{eqnarray}
where $M_{ui} =-\Gamma_{ui} v_i / \sqrt{2}$ and $M_{di} = -\Gamma_{di} v_i/\sqrt{2}$ are all real since spontaneous CP violation is imposed.

The above mass matrices can be transformed into diagonalized form $\hat M_{u,d}$ by bi-unitarity transformation. For Model a), $\hat M_u = \Vckm M_u V_R^{u\dagger}$, and for
Model b), $\hat M_d = \Vckm^\dagger M_d V_R^{d\dagger}$.
A direct identification of the phase $\delsp$ with the phase
$\delta_{\text{KM}}$ in the CKM matrix is not possible in general at this
level. There are, however, classes of mass matrices which allow
such a connection. A simple example is provided by setting $V^{u,d}_R$
to be the unit matrix. With this condition,  one has~\cite{Chen:2007nx}
\begin{eqnarray}
\mbox{Model a)}: &&\Vckm^\dagger = (M_{u1} + e^{i\delsp} M_{u2})\hat M^{-1}_u\nonumber\\
\mbox{Model b)}: &&\Vckm = (M_{d1} + e^{-i\delsp} M_{d2})\hat M^{-1}_d
\label{mvkm}
\end{eqnarray}

Expressing the CKM matrix in the above form is very suggestive. If
$V_{\text{CKM}}$ (or $V_{\text{CKM}}^\dagger$) can always be written as a sum of two
terms with a relative phase, then the phase in the CKM matrix can be
identified with the phase $\delsp$.

Since multi-Higgs bosons are involved, there are in general flavor changing neutral current
(FCNC) interactions mediated by neutral Higgs bosons at the tree
level. However, unlike in general multi-Higgs models, the FCNC Yukawa
couplings are fixed in terms of the quark masses, CKM mixing
angles, and vacuum expectation values $v_i$ of Higgs bosons making phenomenological studies much easier and having different predictions than general
multi-Higgs models~\cite{Chen:2007nx,pheno}. In the following we will study how the same
idea can be applied to lepton sector and related implications.

\section{CP Violating Phase In The lepton sector}

There are some complications for a straightforward generalization of the idea discussed in  the previous section to lepton sector
because neutrinos may be Majorana type.
Also in order to have non-zero neutrino masses in the model, some extensions are needed. We will introduce
right-handed neutrinos to achieve this through seesaw mechanism.

The charged current mixing matrix $\Vpmns$ in the lepton sector is given by,
\begin{eqnarray}
\mathcal{L} = -\frac{g}{\sqrt{2}}\,\bar l_L \gamma^\mu \Vpmns \nu_L W^-_\mu + h.c.
\end{eqnarray}
with $\Vpmns =
V^{l}_L V^{\nu\dagger}_L$. Here $V_L^l$ and $V^\nu_L$ are unitary matrices transforming the charged lepton mass matrix $M_l$ and the neutrino mass matrix $M_\nu$ to their diagonalized form $\hat M_l$ and $\hat M_\nu$,
$M_l =V^{l\dagger}_L  \hat M_l V^l_R$ and $M_\nu = V^{\nu\dagger}_L\hat M_\nu V^{\nu*}_L$.

We consider two possible ways given in the discussion of Ref.~\cite{Chen:2007nx} to couple the leptons to Higgs bosons parallel to the quark sector with PQ charges assigned to leptons as
\begin{eqnarray}
\mbox{Model a)}: &&L_L ( 0)\;, \;\;l_R (-1)\;,\;\;N_R (-1),\nonumber\\
\mbox{Model b)}: &&L_L ( 0)\;, \;\;l_R (+1)\;,\;\;N_R (+ 1)\;.
\end{eqnarray}

Combining with PQ charges for Higgs in Eq.~(\ref{quarkPQ}) the Yukawa interaction of above two models is then given by~\cite{Chen:2007nx}
\begin{eqnarray}
\mbox{Model a)}: &&\mathcal{L} = \overline{ L_L} (Y_1 H_1 + Y_2 H_2 e^{i \delsp}) N_R + \overline{L_L} Y_3
\tilde H_3 l_R + \overline{(N_R)^c} Y_s S  N_R
+ h.c.,\nonumber\\
\mbox{Model b)}: &&\mathcal{L} = \overline{ L_L} Y_3 H_3 N_R + \overline{ L_L} (Y_1 \tilde H_1 + Y_2 \tilde
H_2 e^{-i\delsp}) l_R + \overline{(N_R)^c} Y_s S^\dagger N_R + h.c.\;,
\end{eqnarray}

where $Y_1$, $Y_2$, $Y_3$, and $Y_s$ are $3\times3$ real coupling matrices. The interaction shown above gives the mass terms of lepton sector as~\cite{Chen:2007nx}
\begin{eqnarray}
\mathcal{L}_m = - \overline{ l_L} M_l l_R -
\overline{ \nu_L} M_D N_R - \frac{1}{2} \overline{(N_R)^c} M_R N_R\;,
\end{eqnarray}
with $M_D$, $M_l$, and $M_R$ defined below~\cite{Chen:2007nx}
\begin{eqnarray}
\mbox{Model a)}: &&M_l = - {1\over \sqrt{2}} Y_3 v_3,\;\; M_D = -{1\over
\sqrt{2}}(Y_1 v_1 + Y_2 v_2 e^{i\delsp} )\;,\;\;M_R = - \sqrt{2} Y_s
v_s \;,\nonumber\\
\mbox{Model b)}: &&M_l = - {1\over \sqrt{2}} (Y_1 v_1 + Y_2 v_2 e^{-i\delsp})\;,\;\;
M_D = -{1\over \sqrt{2}} Y_3 v_3,\;\;M_R = -\sqrt{2} Y_s v_s
\;.
\end{eqnarray}

To get neutrino masses, we expressed the neutrino mass terms in the usual seesaw mass matrix with the same notation in Ref.~\cite{He:2009ua} shown as
\begin{eqnarray}
\mathcal{L}^\nu_m=-\frac{1}{2}(\overline{\nu_L},\quad\overline{(N_R)^c})M_{\text{seesaw}}^\dagger
\left
(\begin{array}{c}
(\nu_L)^c\\
N_R\\
\end{array}\right)+h.c.,
\end{eqnarray}
where $M_{\text{seesaw}}$ is defined as $6\times6$ mixing matrix, with
\begin{eqnarray}
M_{\text{seesaw}}=
\left(
\begin{array}{cc}
0&M_D^*\\
M_D^\dagger&M_R\\
\end{array}\right).
\end{eqnarray}

Strictly speaking $V_L^{\nu \dagger}$ is only approximately unitary. The relation between $M_D$ and diagonalized light neutrino mass matrix $\hat{M}_\nu$, also with heavy one $\hat{M}_N=M_R$~\cite{He:2009ua,Schechter:1981cv}, is given by~\cite{Chen:2007nx}
\begin{eqnarray}
M_D \hat{M}^{-1}_N M_D^T=-V_L^{\nu\dagger}\hat{M}_\nu V_L^{\nu *} = (iV_L^{\nu\dagger})\hat{M}_\nu (iV_L^{\nu *}) \label{mDrelation}.
\end{eqnarray}
The general solutions for $M_D$ are given by~\cite{Casas:2001sr}
\begin{eqnarray}
M_D = (iV_L^{\nu\dagger}) \hat{M}_\nu^{1/2}O\hat{M}_N^{1/2}.\label{mD2VmOM}
\end{eqnarray}
where $O$ is a matrix satisfying $O O^T = I$. Choosing $O=I$ is similar to choosing $V_R^d = I$ for Model b) in the quark sector.

We now discuss several simple ways to identify phases in the mass matrix with the spontaneous CP violating phase $\delsp$.

\subsection{Phase in Model a)}

For Model a), we work with the basis where $M_l$ is already diagonalized. If now we redefine $l_L$ to cause $V_L^l$ as $i$, then we have $\Vpmns = i V^{\nu\dagger}_L$ and one can get rid of the `-' sign in the Eq.~(\ref{mD2VmOM}) and set
\begin{eqnarray}
M_D = M_{D1} + M_{D2} e^{i\delsp} = \Vpmns \hat{M}_\nu^{1/2}O\hat{M}_N^{1/2}.
\label{l1}
\end{eqnarray}
where $M_{Di} = - Y_iv_i/\sqrt{2}$ are real matrices. Note that for $l_R$ we also have $V_R^l=i$ at the same time.

In general one can express $M_{D1,2}$ as follows
\begin{eqnarray}
&&M_{D1} = \text{Re}(\Vpmns\hat M_\nu^{1/2} O \hat M_N^{1/2}) -  \cot\delsp \text{Im}(V_{\text{PMNS}}\hat M_\nu^{1/2} O \hat M_N^{1/2})\;,\nonumber\\
&&M_{D2} = {1\over \sin\delsp} \text{Im}(V_{\text{PMNS}}\hat M_\nu^{1/2} O \hat M_N^{1/2})\;.
\end{eqnarray}

There are some complications compared with realization of the same in the quark sector because the Majorana feature of
neutrinos. The $V_{\text{PMNS}}$ matrix can be separated into two parts $\Vpmns =\TVpmns V_p$. Here $\tilde V_{\text{PMNS}}$ is a unitary matrix with only Dirac type of phase. The diagonal phase matrix $V_p$ can be chose by convention to be $V_p = \text{diag}(1, e^{i\sigma_1}, e^{i\sigma_2})$. These phases can
only be measured in $\Delta L = 2$ processes such as neutrinoless double $\beta$ decay. There is also the possibility that phases appear in the matrix $O$.

Identifying $\delsp$ with the phase in $\Vpmns$ is the emphasis of the text. The phase, in general, can appear as the Dirac phase in $\TVpmns$. If the $\TVpmns$ does not have CP violation, for example it is a tri-bimaximal form, then one has to identify $\delsp$ with the phases in $V_p$ or in $O$. We will concentrate on the possibility of
identifying $\delsp$ with the Dirac phase in $\tilde\Vpmns$.

In this case, the spontaneous CP violating phase $\delsp$ is identified with the Dirac phase in $\Vpmns$ which is the only phase in the model.
For the mixing matrix, we can work with the Particle Data
Group (PDG) parametrization of the mixing matrix~\cite{pdg2010}, but with modification to write it in a form with just one phase appearing
with the same sign for simple identification with $\delsp$~\cite{Chen:2007nx}
\begin{eqnarray}
\TVpmns^{\text{PDG}} =
\left( \begin{array}{lll}
c_{12}c_{13}e^{i\delta_{13}}&s_{12}c_{13}e^{i\delta_{13}}&s_{13}\\
-s_{12}c_{23}-c_{12}s_{23}s_{13}e^{i\delta_{13}}&c_{12}c_{23}-s_{12}s_{23}s_{13}
e^{i\delta_{13}}&s_{23}c_{13}\\
s_{12}s_{23}-c_{12}c_{23}s_{13}e^{i\delta_{13}}&-c_{12}s_{23}-s_{12}c_{23}s_{13}
e^{i\delta_{13}}&c_{23}c_{13}
\end{array} \right ),
\end{eqnarray}
where $s_{ij} = \sin\theta_{ij}$ and $c_{ij}=\cos\theta_{ij}$.

The matrices with above parametrization are related to the usual PDG parametrization by multiplying a diagonal phase matrix $V_\delta = \text{diag}(e^{-i\delta_{13}}, 1,1)$ on left-handed side. We make the Dirac phase $\delta_{13}$ in $\Vpmns$ to be identical to $\delsp$, and also because $\delsp$ is identical with Dirac phase in CKM matrix in our model, we take the global fitting results from PDG~\cite{pdg2010} to get $\delta_{13}=(68.9\pm2.7)^\circ$ for CKM matrix with modified PDG parametrization. We have the mass matrices $M_{D1,2}$ given by
\begin{eqnarray}
&&M_{D1}^{\text{PDG}}= \left(\begin{array}{ccc}
0&0&s_{13}\\
-s_{12}c_{23}&c_{12}c_{23}&s_{23}c_{13}\\
s_{12}s_{23}&-c_{12}s_{23}&c_{23}c_{13}\\
\end{array}\right) \hat{M}_\nu^{1/2}O\hat{M}_N^{1/2}\;,\\
&&M_{D2}^{\text{PDG}} = \left(\begin{array}{ccc}
c_{12}c_{13}&s_{12}c_{13}&0\\
-c_{12}s_{23}s_{13}&-s_{12}s_{23}s_{13}&0\\
-c_{12}c_{23}s_{13}&-s_{12}c_{23}s_{13}&0\\
\end{array}\right)\hat{M}_\nu^{1/2}O\hat{M}_N^{1/2}\;.
\end{eqnarray}
The non-zero Dirac $\delta_{13}$ implies none of the mixing angles $\theta_{ij}$ in the $\Vpmns$ can be zero.

Note that the solutions discussed above are not unique.  This implies that in order to completely fix a underlying model, more physical requirements are needed.
To see the non-uniqueness, we take another
parametrization for the mixing matrix, the original
KM matrix~\cite{km}, for illustration. There the mixing matrix is parameterized as
\begin{eqnarray}
\TVpmns^{\text{KM}} =
\left( \begin{array}{lll}
c_1&-s_1c_3&-s_1s_3\\
s_1c_2&c_1c_2c_3-s_2s_3e^{i\delta}&c_1c_2s_3+s_2c_3e^{i\delta}\\
s_1s_2&c_1s_2c_3+c_2s_3e^{i\delta}&c_1s_2s_3-c_2c_3e^{i\delta}
\end{array} \right ),\label{KMparametrization}
\end{eqnarray}
with $s_{i} = \sin\theta_{i}$ and $c_{i}=\cos\theta_{i}$. In this case one identifies $\delta=\delsp$ with $\delta = (90.3\pm2.7)^{\circ}$ obtained from PDG global fitting results~\cite{pdg2010}, and the mass matrices $M_{D1,2}$ can be written as
\begin{eqnarray}
&&M_{D1}^{\text{KM}}= \left(\begin{array}{ccc}
c_1&-s_1c_3&-s_1s_3\\
s_1c_2&c_1c_2c_3&c_1c_2s_3\\
s_1s_2&c_1s_2c_3&c_1s_2s_3\\
\end{array}\right) \hat{M}_\nu^{1/2}O\hat{M}_N^{1/2}\;,\\
&&M_{D2}^{\text{KM}} =
\left(\begin{array}{ccc}
0&0&0\\
0&-s_2s_3&s_2c_3\\
0&c_2s_3&-c_2c_3\\
\end{array}\right)
\hat{M}_\nu^{1/2}O\hat{M}_N^{1/2}\;.
\end{eqnarray}

We emphases that the above two ways of parameterizing the mixing matrices are two different models. We will refer them as Model a(PDG)) and Model a(KM)). The differences between these two models will show up in the CP violating Jarlskog parameter~\cite{Jarlskog} $J$, in neutrinoless double beta decay and also in the Yukawa couplings since $M_{Di}$ for these two models are different.

Since we have
identified the Dirac phase with the spontaneous CP violating phase which are also related to the quark mixing CP violating phase, the Dirac phase in $\tilde \Vpmns$ is known. For Model a(PDG)), using the
current range of mixing parameters determined from various neutrino oscillation data~\cite{GonzalezGarcia:2010er}
\begin{eqnarray}
\theta_{12}=34.5\pm1.0\left(^{+3.2}_{-2.8}\right)^\circ,\;
\theta_{23}=42.8^{+4.7}_{-2.9}\left(^{+10.7}_{-7.3}\right)^\circ,\;
\theta_{13}=5.1^{+3.0}_{-3.3}(\leq12.0)^{\circ},\label{PDGparametrizationExp}
\end{eqnarray}
we obtain the central and 1$\sigma$ allowed range for $J(\text{PDG})$
\begin{eqnarray}
J(\text{PDG})=0.019\pm0.012.
\end{eqnarray}

For Model a(KM)), we can obtain the central and $1\sigma$ ranges for the mixing angles from $|V_{e1}|$, $|V_{e2}|$, and $|V_{e3}|$ by using the values in Eq.~(\ref{PDGparametrizationExp}). The results are
\begin{eqnarray}
\sin\theta_{1}=0.57\pm0.02;\;\;\sin\theta_{3}=0.16\pm0.10.\label{KMs1s3}
\end{eqnarray}
Since with this model one also identifies the phase in Eq.~(\ref{KMparametrization}) with Dirac phase in the original KM parametrization of quark sector, that is, $\delta=(90.3\pm2.7)^{\circ}$ as shown before, Using this value with Eq.~(\ref{KMs1s3}) we can derived the corresponding values for $s_2$ from $|V_{\mu3}|$ or $|V_{\tau3}|$, which is given by
\begin{eqnarray}
\sin\theta_{2}=0.68\pm0.04.
\end{eqnarray}

The value for Jarlskog parameter in this case $J(\text{KM})$ is different than that for $J(\text{PDG})$. Using the above derived values for mixing angles and phase, we have
\begin{eqnarray}
J(\text{KM}) = 0.021\pm0.013.
\end{eqnarray}
In principle measurement of $J$ can be used to distinguish different models, but the difference is small to it practically difficult.

The above two models also have difference in predicting the effective mass $<m_{\beta\beta}> = |\sum_i m_i V^2_{ei}|$ for neutrinoless double beta decay which we discuss in the following.

At present, the absolute mass scale for neutrino is not known although stringent constraint exists from cosmological consideration, which gives~\cite{DeBernardis:2008qq} $\sum_i m_i<0.28\text{eV}$. Combining with data from neutrino oscillation
\begin{eqnarray}
\Delta m_{12}^2&=&(7.59\pm0.20)\times10^{-5}(\text{eV})^2\text{\cite{Aharmim:2008kc}};\\
\Delta m_{23}^2&=&(2.43\pm0.13)\times10^{-3}(\text{eV})^2\text{\cite{Adamson:2008zt}},
\end{eqnarray}
one can constrain $<m_{\beta\beta}>$.

\begin{figure}
\includegraphics[width=11cm]{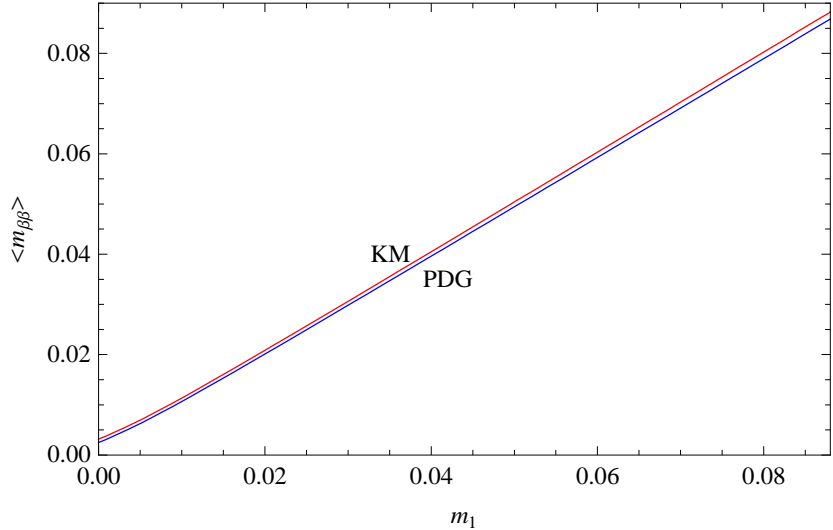}
\caption{$< m_{\beta\beta}>$ vs. $m_1$ for model $a)$, with normal hierarchy, respectively. }\label{CCplot}
\end{figure}

We will take the central values of the mixing angles and phases, neutrino mass squared differences, and also take neutrino mass $m_1$ as a free parameter satisfying the cosmological constraint for illustration. The results are show in Figure \ref{CCplot}. We have only plotted the case with normal neutrino mass hierarchy. The results with inverted hierarchy are almost the same and have not been plotted. As can be seen from the figure that the two different models lead to similar results. There the $<m_{\beta\beta}>$ are in principle different but the differences are small.

Another quantity related to neutrino mass is $<m_{\nu_e}> = (\sum_i m_i^2 |V_{ei}|^2)^{1/2}$ which can be measured in tritium decay. This quantity is however, the same for the two models we considered above. With the central values for mixing angles, it is given by 0.088 for normal hierarchy with $m_1 = 0.088$ close to its upper bound, for example. For inverted hierarchy with the same $m_1$, we get almost the same results as that we have in normal hierarchy. This quantity is not useful in distinguishing different models considered above.

\subsection{ Phase in Model b)}

For Model b), information about CP violation is encoded in $M_l$. $M_D$ is a real matrix from Yukawa coupling $Y_3$. If one recalls Eq.~(\ref{mD2VmOM}), it is straightforward to deduce that $V_L^\nu$ must be complex. For simplicity we will work in a basis with $V_L^\nu=i$ and we have a simpler relation
\begin{eqnarray}
M_D=\hat{M}_\nu^{1/2} O \hat{M}_N^{1/2},
\end{eqnarray}
where $O$ is a complex orthogonal matrix as mentioned before. we then follow the same philosophy to have set a possible non-unit unitary matrix $V_R^l$ to $i$ in order to getting the useful relation which is given as
\begin{eqnarray}
M_l = M_{l1} + M_{l2} e^{-i\delsp} = \Vpmns^\dagger \hat M_l\;.
\label{l2}
\end{eqnarray}
In this basis $\Vpmns =-iV_L^l$, and we can apply the same idea in model(a) that Dirac phase $\Vpmns$ is identical to $\delsp$. We also work with two models with the PDG and original KM parameterizations, Model b(PDG)) and Model b(KM)). For Model b(PDG)), we have
\begin{eqnarray}
&&M_{l1}^{\text{PDG}} =\left(\begin{array}{ccc}
0&-s_{12}c_{23}&s_{12}s_{23}\\
0&c_{12}c_{23}&-c_{12}s_{23}\\
s_{13}&s_{23}c_{13}&c_{23}c_{13}\\
\end{array}\right)\hat M_l,M_{l2}^{\text{PDG}}= \left(\begin{array}{ccc}
c_{12}c_{13}&-c_{12}s_{23}s_{13}&-c_{12}c_{23}s_{13}\\
s_{12}c_{13}&-s_{12}s_{23}s_{13}&-s_{12}c_{23}s_{13}\\
0&0&0\\
\end{array}\right)\hat{M}_l.\nonumber\\
\end{eqnarray}

For Model b(KM)), we have
\begin{eqnarray}
&&M_{l1}^{\text{KM}} =\left(\begin{array}{ccc}
c_1&s_1c_2&s_1s_2\\
-s_1c_3&c_1c_2c_3&c_1s_2c_3\\
-s_1s_3&c_1c_2s_3&c_1s_2s_3\\
\end{array}\right)\hat M_l\;,\;\;\;M_{l2}^{\text{KM}}= \left(\begin{array}{ccc}
0&0&0\\
0&-s_2s_3&c_2s_3\\
0&s_2c_3&-c_2c_3\\
\end{array}\right)\hat{M}_l\;.
\end{eqnarray}

The Jarlskog parameters $J$ and the effective mass $<m_{\beta\beta}>$ in the two models are the same as those in models a(PDG)) and b(KM)).

\section{Neutral Higgs and charged lepton Yukawa couplings}

The different ways to identify the CP violating phase in the lepton sector with that from the spontaneous symmetry breaking sector also restrict the forms of the Yukawa couplings in the model differently. New interaction due to Higgs exchange can generate some interesting phenomena.
Here we display the neutral Higgs coupling to the charged leptons for Models a) and b) and discuss some consequences.
We have

\begin{eqnarray}
&&\mbox{Model a)}:\;\;
\mathcal{L}_{llh^0}=-\bar{l}_L \hat{M}_l l_R\bigg[\frac{v_{12}^2v_s}{N_Av_3}(H_2^0-ia_2)
+\frac{1}{v}H_3^0+\frac{v_{12}^2}{N_a}(H_4^0-ia)\bigg]+h.c.\;,\\
&&\mbox{Model b)}:\;\;
\mathcal{L}_{llh^0}=-\overline{l_{L}}\bigg(-\frac{v_1}{v_2v_{12}}\hat{M}_l
+\frac{v_{12}}{v_1v_2}\Vpmns M_{l1}\bigg)l_{R}(H_1^0-ia_1)\nonumber\\
&&\hspace{3.5cm}+\overline{l_{L}}\hat{M}_l l_{R}\bigg[\frac{v_3v_s}{N_A}(H_2^0-ia_2)
-\frac{1}{v}H_3^0+\frac{v_3^2}{N_a}(H_4^0-ia)\bigg]+h.c.,
\end{eqnarray}
where $l_{R,L}$ shown in above two equations are charged lepton mass eigenstates. Other Yukawa couplings are given in the Appendix.

It is interesting to note that for Model a), the couplings are flavor conserving, but for Model b) the couplings can have neutral flavor changing current at the tree level. This can be used to distinguish these two models by looking at, for example $\mu \to e e \bar e$ and $\tau \to l_1 l_2\bar l_3$ decays. These decays have not been observed, but there are stringent bounds from various experimental measurements with
\begin{eqnarray}
&&\text{Br}(\tau\to e e\bar{e})_{\text{exp}}<3.6 \times10^{-8}\;\text{\cite{Miyazaki:2007zw}},\;\;
\text{Br}(\tau\to \mu \mu \bar{\mu})_{\text{exp}}< 3.2 \times10^{-8}\;\text{\cite{Miyazaki:2007zw}},\nonumber\\
&&\text{Br}(\tau\to \mu e\bar{e})_{\text{exp}}<2.7 \times10^{-8}\;\text{\cite{Miyazaki:2007zw}},\;\;
\text{Br}(\tau\to e \mu \bar{\mu})_{\text{exp}}<3.7 \times10^{-8}\;\text{\cite{Aubert:2007pw}},\nonumber\\
&&\text{Br}(\tau\to \mu\mu\bar{e})_{\text{exp}}<2.3\times10^{-8}\;\text{\cite{Miyazaki:2007zw}},\;\;
\text{Br}(\tau\to  e  e \bar{\mu})_{\text{exp}}< 2.0 \times10^{-8}\;\text{\cite{Miyazaki:2007zw}},\nonumber\\
&&\text{Br}(\mu\to ee\bar{e})_{\text{exp}}<1.0\times10^{-12}\;\text{\cite{Bellgardt:1987du}}.\label{decay}
\end{eqnarray}

In the following we study these decay modes in more details to see if it is also possible to distinguish the Model b(PDG)) and Model b(KM)).
We have
\begin{eqnarray}
\Vpmns^{\text{PDG}} M_{l1}^{\text{PDG}}&=&\left(
\begin{array}{ccc}
s_{13}^2&s_{23}s_{13}c_{13}&c_{23}s_{13}c_{13}\\
s_{23}s_{13}c_{13}&c_{23}^2+s_{23}^2c_{13}^2&-s_{23}c_{23}s_{13}^2\\
c_{23}c_{13}s_{13}&-s_{23}c_{23}s_{13}^2&s_{23}^2+c_{23}^2c_{13}^2\\
\end{array}
\right)\hat{M}_l,\;\\
\Vpmns^{\text{KM}} M_{l1}^{\text{KM}}&=&\left(
\begin{array}{ccc}
1&0&0\\
0&c_2^2&c_2s_2\\
0&c_2s_2&s_2^2\\
\end{array}
\right)\hat{M}_l.
\end{eqnarray}

\begin{figure}[h!]
\includegraphics[width=3in]{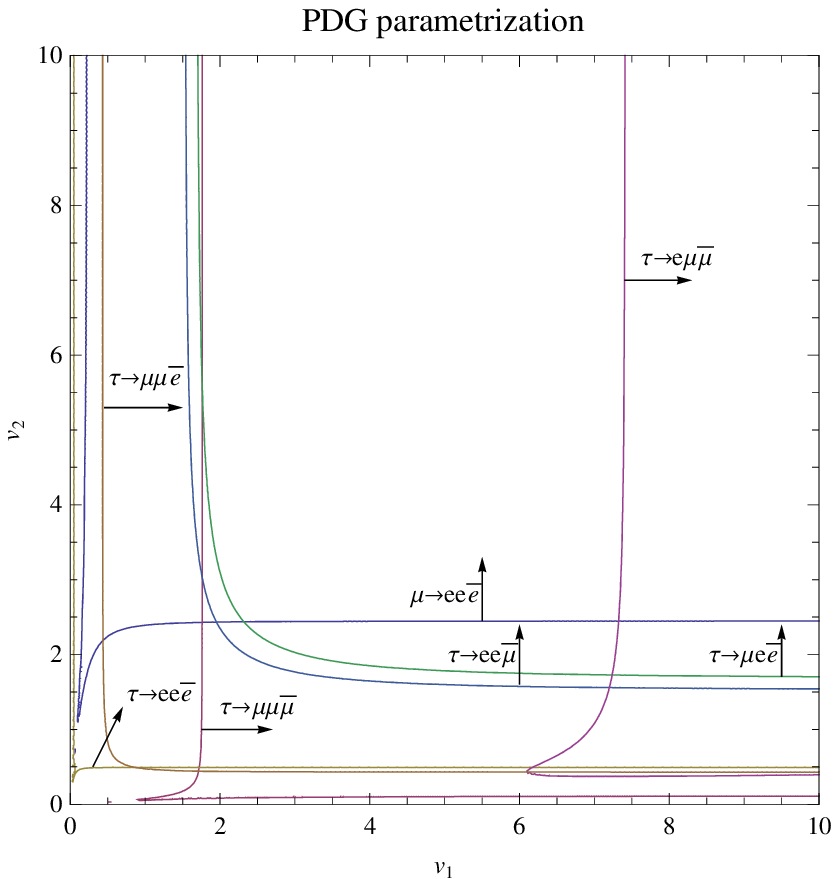}\hspace{1cm}\includegraphics[width=3in]{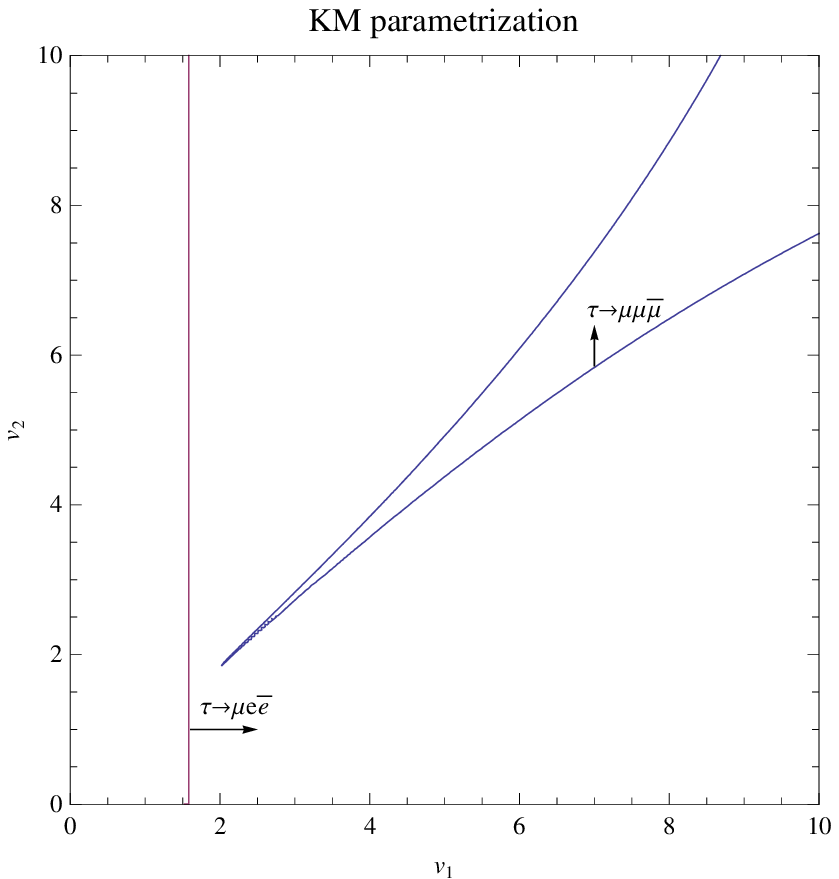}
\caption{The contour plots of $v_1$ and $v_2$ in the range of 1 to 10 GeV for the upper bounds of $\mu \to e e\bar e$, and $\tau \to l_1 l_2 \bar l_3$. Here
$l_i$ can be one of the e or $\mu$. The directions pointed by the arrows are regions satisfying the upper bounds.}\label{Brplot}
\end{figure}

These two models have different predictions. For Model b(PDG)), at the tree level it is possible to have all the 7 decay modes listed in Eq.~(\ref{decay}), but for Model b(KM)) it only allows $\tau \to \mu \mu \bar{\mu}$ and  $\tau \to \mu e \bar{e}$ to happen at tree level.
The decay rates of these processes depend on the PMNS mixing angles and phase, the Higgs masses, and also the two VEV's $v_1$ and $v_2$. To have some concrete feeling about how experimental data put constraints on the models, we take the PMNS parameters with their central values described earlier, and the Higgs mass mediating FCNC interaction to be 100 GeV for illustrations. With these parameters fixed, the branching ratios depend on $v_1$ and $v_2$. Given the upper bound of a branching ratio, one can finds the allowed regions of $v_{1,2}$.

We show the results in Fig.~2. We see, from the figure for Model b(PDG)), that the combined upper bounds $\mu \to e e \bar e$ and $\tau \to e \mu \bar \mu$ give the most severe constraints requiring $v_1$ to be larger than 7.4 GeV and $v_2$ to be larger than 2.5 GeV. Once these are satisfied, the other bounds are satisfied. For Model b(KM)), $\tau \to \mu\mu \bar \mu$ gives the strongest constraint. The allowed region for $v_{1,2}$ for Model b(PDG) is bigger than that for Model b(KM)). For Model b(PDG)) it is possible to have $\mu \to e e \bar e$ to be close to its upper bound, but for Model b(KM)) it is zero. Future experimental data can distinguish these two types of models.

\section{Conclusion}
We have studied some implications of models for identifying the spontaneous CP violating phase to the Dirac phase in PMNS matrix. This identification forces the Yukawa couplings for lepton sector only dependent on lepton masses,  mixing angles in PMNS matrix, and VEV's of Higgs bosons. Using different parameterizations of the PMNS matrix results in different models.  We have studied in detail two popular, the PDG and the original KM, parameterizations. We then studied effects on properties of the PMNS matrix and neutrino masses. Since the CP violating phase in the CKM mixing matrix is determined by experimental data, CP violating phase in the lepton sector is also fixed. The mass matrix for neutrinos is constrained leading to constraints on the Jarlskog CP violating parameter $J$, and the effective mass $<m_{\beta\beta}>$ for neutrinoless double beta decay. The Yukawa couplings are also constrained. Different ways of identifying the phases have different predictions for $\mu \to e e\bar e$ and $\tau \to l_1
  l_2 \bar l_3$. Future experimental data can be used to distinguish different models.

\noindent {\bf Acknowledgments}$\,$ This work was supported in
part by the 985 project, NSC and NCTS.

\begin{appendix}
\section{}
To obtain the Yukawa couplings, it is best to work in the basis where un-physical Higgs have been removed.
The un-physical Higgs bosons are the Goldstone fields $h_w$ and $h_z$ ``eaten'' by $W$
and $Z$. There is also the axion field $a$ which is invisible in this model. They are~\cite{Chen:2007nx}
\begin{eqnarray}
&&h_w = {1\over v}(v_1 h^-_1 + v_2 h^-_2 + v_3
h^-_3)\;,\nonumber\\
&&h_z = {1\over v}(v_1 A_1 + v_2 A_2 + v_3 A_3)\;,\nonumber\\
&& a = (-v_1 v^2_3 A_1 -v_2 v^2_3 A_2 +  v^2_{12}v_3 A_3 -  v^2
v_s A_s)/N_a\;,
\end{eqnarray}
where $v^2 = v^2_1+v^2_2 +v^2_3$ and $N_a^2 = v^2(v_{12}^2v_3^2 +
v_s^2v^2 )$ with $v^2_{12} = v^2_1 +v^2_2$.

The physical fields, $a_{1,2}$, $a$ and $H^0_i$ related to the original fields are given by~\cite{Chen:2007nx}
\begin{eqnarray}\label{mixing}
&&\left (\begin{array}{c}A_1\\A_2\\A_3\\A_s\end{array}\right ) =
\left ( \begin{array}{cccc} v_2/v_{12}&-v_1v_3
v_s/N_A&v_1/v&-v_1v_3^2/N_a\\
-v_1/v_{12}&-v_2v_3
v_s/N_A&v_2/v&-v_2v_3^2/N_a\\
0&v^2_{12}
v_s/N_A&v_3/v&v_{12}^2v_3/N_a\\
0&v^2_{12} v_3/N_A&0&-v^2v_s/N_a\end{array}\right
)\left ( \begin{array}{c}a_1\\a_2\\h_z\\a \end{array}\right )\;,
\nonumber\\
&&\left (\begin{array}{c}h^-_1\\h^-_2\\h^-_3\end{array}\right ) =
\left ( \begin{array}{ccc} v_2/v_{12}&v_1v_3/v v_{12}&v_1/v\\
-v_1/v_{12}&v_2v_3
/v v_{12}&v_2/v\\
0&-v_{12}/v&v_3/v\end{array}\right )\left (
\begin{array}{c}H^-_1\\H^-_2\\h_w \end{array}\right )\;,
\end{eqnarray}
where $N_A^2 = v^2_{12}(v^2_{12}v^2_3 +v_s^2v^2)$. $a_{1,2}$ and
$H^-_{1,2}$ are the physical degrees of freedom for the Higgs
fields. With the same rotation as that for the neutral pseudoscalar,
the neutral scalar Higgs fields $(R_1,R_2,R_3,R_s)^T$ become
$(H_1^0,H_2^0,H^0_3, H^0_4)^T$. Since the invisible axion scale $v_s$
is much larger than the electroweak scale, to a very good
approximation, $N_a = v^2v_s$ and $N_A = v_{12} v v_s$.

Rewriting the Yukawa couplings in terms of the Higgs fields shown above, $H_i^0$, $a_i$ and axion $a$, we have
\begin{eqnarray}
\mathcal{L}_{\nu l h^\pm}
=&-&i\sqrt{2}\;\overline{l_L}\left(-\frac{v_1}{v_2v_{12}}M_D
+\frac{v_{12}}{v_1v_2}M_{D1}\right)N_RH_1^-\nonumber\\
&-&i\sqrt{2}\left(\;\overline{l_L}\frac{v_3}{vv_{12}}M_D N_R
+\;\overline{l_R}\frac{v_{12}}{vv_3}\hat{M}_l\nu_{L}\right)H_2^-+h.c..\label{LmaHc}
\end{eqnarray}
The couplings of neutral Higgs with neutrinos are given by
\begin{eqnarray}
\mathcal{L}_{\nu \nu h^0}
=&-&\overline{\nu_L}\bigg[\left(-M_D\frac{v_1}{v_2v_{12}}+M_{D1}\frac{v_{12}}{v_1v_2}\right)(H_1^0+ia_1)
-\frac{v_3v_s}{N_A}M_D(H_2^0+ia_2) \nonumber\\
&+&\frac{M_D}{v}H_3^0-\frac{v_3^2M_D}{N_a}(H_4^0+ia) \bigg]N_R
-\frac{1}{2}\overline{(N_R)^c}\hat{M}_N N_R\bigg[\frac{v_{12}^2v_3}{v_sN_A}(H_2^0+ia_2)\nonumber\\
& -& \frac{v^2}{N_a}(H_4^0+ia)\bigg]+h.c..
\end{eqnarray}

For Model b), Yukawa couplings of charged Higgs and neutral Higgs with neutrinos are shown as
\begin{eqnarray}
\mathcal{L}_{\nu l h^\pm}&=&\sqrt{2}\,i\,\overline{l_{L}}\frac{v_{12}}{vv_3}\Vpmns M_D N_RH_2^-
-\sqrt{2}\,i\overline{\nu_{L}}\left(-\frac{v_1}{v_2v_{12}} M_l
+\frac{v_{12}}{v_1v_2}  M_{l1}\right)l_{R}H_1^+\nonumber\\
&-&\sqrt{2}\,i\overline{\nu_{L}}\frac{v_3}{vv_{12}}M_ll_{R}H_2^++h.c.,\\
\
\mathcal{L}_{\nu \bar \nu h^0}&=&-\overline{\nu_L}M_D N_R\bigg[\frac{v_{12}^2v_s}{N_A v_3}(H_2^0+ia_2)+\frac{1}{v}H_3^0
+\frac{v_{12}^2}{N_a}(H_4^0+ia)\bigg]\nonumber\\
&-&\frac{1}{2}\overline{(N_R)^c}\hat{M}_N N_R\bigg[\frac{v_{12}^2v_3}{N_A v_s}(H_2^0-ia_2)
-\frac{v^2}{N_a}(H_4^0-ia)\bigg].\label{LmbHn}
\end{eqnarray}
\end{appendix}
Note that in Eq.~(\ref{LmaHc}-\ref{LmbHn}) $l_L$ and $l_R$ are already as their mass eigenstates. The neutral Higgs couplings to charged leptons have been given in the text.

\end{document}